\newif\ifabstract
\abstracttrue
 \abstractfalse 
\newif\iffull
\ifabstract \fullfalse \else \fulltrue \fi

\documentclass[11pt]{article}
\usepackage{amsfonts}
\usepackage{amssymb}
\usepackage{amstext}
\usepackage{amsmath}
\usepackage{xspace}
\usepackage{theorem}
\usepackage{graphicx}
\usepackage{url}
\usepackage{graphics}
\usepackage{colordvi}
\usepackage{colordvi}
\usepackage{subfigure}
\usepackage{hyperref}

\advance \oddsidemargin by -0.8in
\newcommand{\myparskip}{3pt}
\parskip \myparskip

\begin{document}

\title{A Fault-Tolerant Distributed Detection of Two Simultaneous Events in Wireless Sensor Networks}
\date{}

\author{Mina Moradi Kordmahalleh\thanks{Department of Electrical and Computer Engineering,
North Carolina A\&T State University. Email: {\tt mmoradik@aggies.ncat.edu}}\and {Mohammad Gorji Sefidmazgi\thanks{School of Information, The University of Arizona. Email: {\tt mgorjise@email.arizona.edu}}} \and Jafar Ghaisari\thanks{Department of Electrical and Computer Engineering, Isfahan University of Technology. Email: {\tt \{ghaisari,j-askari\}@cc.iut.ac.ir}} \and Javad Askari\footnotemark[3]}

\maketitle

\thispagestyle{empty}

\begin{abstract}
Wireless Sensor Networks (WSNs) consist of many low cost and light sensors dispersed in an area to monitor the physical environment. Event detection in WSN area, especially detection of multi-events at the same time, is an important problem.  This article is a new attempt for detection of two simultaneous events based on distributed data processing structure and Bayesian criteria.  For accurate detection of two simultaneous events, we proposed new decision rules based on likelihood ratio test and also derived probability of detection error based on Bayesian criteria. In addition to multi-event detection, the proposed method is expanded to a fault-tolerant procedure if there are faults in decision making of sensors.  Performance of the proposed approach is demonstrated for detection of events in different circumstances. Results show the effectiveness of the algorithm for fault-tolerant multi-event detection.
\end{abstract}


\section{Introduction}\label{sec: intro}
A Wireless Sensor Network (WSN) comprises many small sensor nodes distributed in an area.  Each sensor node consists of measurement devices, computational resources, wireless communication components, and a finite power source \cite{Akyildiz2002,Zhao2004}.  The cooperation, flexibility and ability of sensor nodes to work in diverse environments create opportunities for widespread employment of WSNs in different applications such as healthcare, environment, agriculture, military and industry \cite{Pottie2000,Szewczyk2004,Aslan2012,khan2016wireless}.

One of the most important applications of WSN is detection of an event based on raw data of the sensor nodes. An event is defined as an unusual change in the environmental parameters such as temperature, humidity, pressure \cite{Wang2008,Aslan2012,Diaz-Ramirez2012}. In \textit{centralized detection}, the raw data sampled by the sensors are transmitted to a base station which has more storage and computation power than the sensor nodes.   In \textit{distributed detection}, the raw data is processed in each sensor node and then the obtained results from neighboring sensors are processed in a higher level. Due to limited energy resources and communication bandwidth, the distributed detection methods have received more attention \cite{Blum1997,Wang2005,mina}. 

Due to sensor malfunction or harsh environments, sensor nodes might fail and provide faulty measurements which lead to wrong detection alarms \cite{Chen2006,Mahapatro2013}.  For instance, a faulty sensor node may report an event while no event happened.  In order to ensure the detection accuracy, several fault-tolerant detection algorithms were introduced in literature.  A fault-tolerant event detection method based on distributed Bayesian algorithm was proposed in \cite{Krishnamachari2004}.  A two-level event detection using neural networks and Bayesian classifier was proposed in \cite{Bahrepour2009}. A distributed fault identification algorithm was introduced in \cite{Ding2005} in which sensor nodes compare their own measurements with the median of neighbors' data to detect the event region.  In \cite{Yim2010}, a distributed adaptive fault-tolerant event detection  used confidence levels of sensor nodes to adjust the threshold of decision rules.  In \cite{Ping2009}, the weighted distances between sensor nodes and event region was utilized for distributed detection.  In \cite{Luo2006}, an energy-efficient distributed fault-tolerant method based on Bayesian and Neyman-Pearson approaches was introduced.  A grid-based distributed event detection scheme for WSN was introduced in \cite{Ko2011}.  In \cite{Chen2016108}, a method based on Markov Random field was proposed which used the spatial and temporal correlation of sensor data for event detection. Detection problem in \cite{7145465} was formulated  as a binary hypothesis testing and optimal decision rules were designed using the Poisson Point Process and Binomial Point Process. 

Aforementioned methods are only applicable for detection of a single event in WSN area.  However, more than one event may occur, especially in the pollution monitoring and fire detection applications. Although detection of simultaneous events is an important field of study, but there are limited number of researches in this area. In \cite{Banerjee2008}, a multiple event detection method was proposed that used the correlation among sensor outputs to find the region of multiple events.  The methods of \cite{Liu2011,6881722} have used the assumption that the region of events are sparse, and then converted the event detection to $\ell_1$ regularized least squares problem.

In this paper, our focus is on detection of two different events which happened simultaneously in distinct regions of WSN area.  The proposed  detection procedure is a two-layer distributed detection algorithm where only data from neighboring sensor nodes are used for decision making.  This implementation decreased the load of communication among sensor nodes. For  detection of two events, we proposed the decision rules based on a set of likelihood ratio tests.  Using the decision rules and Bayesian criteria, we formulated the probability of detection error of two simultaneous events.  To have a fault-tolerant detection algorithm, we defined different types of sensor faults in decision making. By considering the probability of sensor faults, the decision rules and probability of detection error for two simultaneous events are modified.
%
%

The rest of the paper is organized as follows.  In Section \ref{Sec:2.1}, detection of two simultaneous events in WSNs and the corresponding decision rules are formulated based on our proposed distributed method.  Section \ref{Sec:2.2} defines the decision metric criteria which are used in the obtained probability of detection error.  In Section \ref{Sec:2.3}, the algorithm is modified for the case of decision making in presence of sensor faults.  Section \ref{Sec:3} illustrates different examples and simulation results to verify the performance of the proposed method for detection of two simultaneous events in WSN area.


\section{Distributed Detection of Two Simultaneous Events} \label{Sec:2}

\subsection{Decision Rules} \label{Sec:2.1}
\begin{figure}
\centering
\includegraphics[scale=1.5]{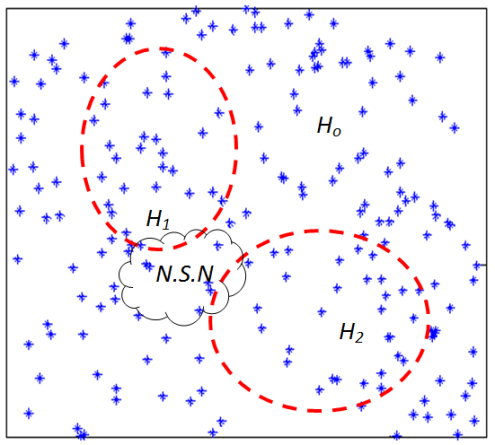}
\caption{A sample area of WSN contains $N$ sensor nodes (blue points). A NSN includes the neighboring nodes inside the communication range of each sensor node. The regions of two distinct simultaneous  events  are show with red lines. }
\label{f1}
\end{figure}

Suppose that $N$ sensor nodes are scattered all over the interested region of a WSN.  Each sensor node can communicate with its own \textit{Neighboring Sensor Nodes (NSN)} in its communication range.  Assume that two events (\textit{event1} and \textit{event2}) with different statistical properties than the normal situation are occurred simultaneously in distinct regions of the WSN. The events and normal situation are represented by three hypotheses $H0$, $H1$ and $H2$ in Eq. \ref{eq1}. Fig. \ref{f1} includes a sample WSN area where the regions of two events are  shown by dashed lines. 

\begin{equation}
\label{eq1}
\begin{cases}
H0&\text{:\textit{Normal (No event)}}   \\ 
H1&\text{:\textit{event1}}  \\ 
H2&\text{:\textit{event2}}  
\end{cases}
\end{equation}

\begin{figure}[t]
\centering
\includegraphics[scale=.8]{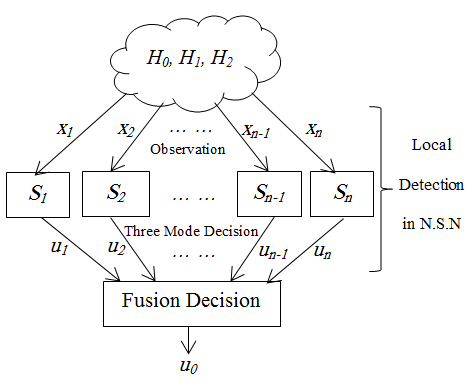}
\caption{In distributed two-layer detection of two simultaneous events, the final three mode decision $u_0 \in\{0,1,-1\}$ at each sensor node is obtained using the local decisions of $n$ neighboring sensor nodes $S_i,  i=1,\dots,n$.}
\label{f2}
\end{figure}

Our goal is to find the decision rules to determine the event that happened in each sensor node.  To this aim, a distributed two-layer detection algorithm is proposed which is shown in Fig. \ref{f2}. 
At each sensor node $j$ where $j=1,\dots,N$, the local decision $u_j\in\{0,1,-1\}$ is made based on noisy observation $x_j$ of the sensor node. Here, decisions 0, 1, -1 represent the normal, \textit{event1} and \textit{event2} respectively. Suppose that each sensor node can communicate with $n$ neighboring nodes. To make the final decision $u_0$ at each sensor node, the local decisions $u_i$  where $i=1,\dots,n$  are used in our modified version of $k$ out of $n$ rule \cite{Clouqueur2004}. This procedure is performed in every sensor node in the WSN.

In a case that the hypothesis $H_j, j=0, 1, 2$ is true, $x_i, i=1,\dots, N$ follows the probability distribution function ${P_{{x_i}|{H_j}}}(X|{H_j})$.  For local decision making in each sensor node, two ratios $\Lambda_1$ and $\Lambda_2$ are represented as:
\begin{align}
{\Lambda _1}(X) \buildrel \Delta \over = \frac{{{P_{{x_i}|{H_1}}}(X|{H_1})}}{{{P_{{x_i}|{H_0}}}(X|{H_0})}}\\
{\Lambda _2}(X) \buildrel \Delta \over = \frac{{{P_{{x_i}|{H_2}}}(X|{H_2})}}{{{P_{{x_i}|{H_0}}}(X|{H_0})}}
\end{align}
Eqs. \ref{eq:4},\ref{eq:5} and \ref{eq:6} are likelihood ratio tests for local decision making in each sensor node. Here, $\lambda_1$ and $\lambda_2$ are thresholds of comparison. For example in Eq. \ref{eq:4}, if $\Lambda_1 >  \lambda_1 $, then hypothesis $H_1$ or $H_2$ are decided. Otherwise, $H_0$ or $H_2$ are decided. 

\begin{align}
{\Lambda _1}(X)\frac{{\mathop  > \limits^{{H_1} {\rm{or}} {H_2}} }}{{\mathop  < \limits_{{H_0} {\rm{or}} {H_2}} }}{\lambda _1}\label{eq:4}\\
{\Lambda _2}(X)\frac{{\mathop  > \limits^{{H_2} {\rm{or}} {H_1}} }}{{\mathop  < \limits_{{H_0} {\rm{or}} {H_1}} }}{\lambda _2}\label{eq:5}\\
\frac{{{\Lambda _2}(X)}}{{{\Lambda _1}(X)}}\frac{{\mathop  > \limits^{{H_2} {\rm{or}} {H_0}} }}{{\mathop  < \limits_{{H_1} {\rm{or}} {H_0}} }}\frac{{{\lambda _2}}}{{{\lambda _1}}}\label{eq:6}
\end{align}
Intersection of these tests determines the local decision which is summarized in a simplified form.   

\begin{align}
\label{eq:7}
\begin{cases}
\text{If } & \Lambda _1(X) < \lambda _1 \text{ and }  \Lambda _2(X) < \lambda _2 \text{ then } H_0 \text{ is selected.}\\
\text{If } & \Lambda _1(X) > \lambda _1 \text{ and }    \frac{\Lambda _2(X)}{\Lambda _1(X)}  < \frac{\lambda _2(X)}{\lambda _1(X)} \text{ then } H_1 \text{ is selected.}\\
\text{If } & \Lambda _2(X) > \lambda _2 \text{ and }    \frac{\Lambda _2(X)}{\Lambda _1(X)}  > \frac{\lambda _2(X)}{\lambda _1(X)} \text{ then } H_2 \text{ is selected.}
\end{cases}
\end{align}

If one of the $H_0$, $H_1$ or $H_2$ is selected for a sensor node, then the local decision $u_i$ is declared as 0, +1, or -1 respectively.  After determining the local decisions for all of $n$ neighboring sensors, final decision in each sensor node $u_0$ is made in the second detection layer by the \textit{Modified $k$ out of $n$ rule}. If $k$ out of $n$ local decisions in neighboring sensors are equal to 1 (or -1), then the final decision in the sensor node is 1 (or -1) meaning that the first event (or second event) is decided. Otherwise, the final decision is equal to 0 means that no event happened on that sensor node.

In order to derive the decision rules based on the sensor observations $x_i$, let's consider that $x_i$ is sampled from a Gaussian distribution with unit variance and mean $m \in \{ m_0, m_1, m_2 \}$ for normal, \textit{event1} and \textit{event2} respectively. Then, the hypotheses are defined over $x_i$:


\begin{equation}
\begin{cases}
H0 & \text{: $\mathcal{N}(m_0,1)$ \textit{  No event} }   \\ 
H1 & \text{: $\mathcal{N}(m_1,1)$ \textit{  event1}  }  \\ 
H2 & \text{: $\mathcal{N}(m_2,1)$ \textit{  event2}   }  
\end{cases}
\end{equation}
Without losing of generality, by assuming that $m_2>m_1>m_0$, the likelihood tests for local decision making in Eqs. \ref{eq:4},\ref{eq:5} and \ref{eq:6} are reformulated as:
\begin{align}
\begin{cases}
{x_i}\frac{{\mathop  > \limits^{^{{H_1} {\rm{or}} {H_2}}} }}{{\mathop  < \limits_{_{{H_0} {\rm{or}} {H_2}}} }}\frac{1}{{({m_1} - {m_0})}}\ln ({\lambda _1}) + \frac{{({m_1} + {m_0})}}{2}\\
{x_i}\frac{{\mathop  > \limits^{^{{H_1} {\rm{or}} {H_2}}} }}{{\mathop  < \limits_{_{{H_0} {\rm{or}} {H_1}}} }}\frac{1}{{({m_2} - {m_0})}}\ln ({\lambda _2}) + \frac{{({m_2} + {m_0})}}{2}\\
{x_i}\frac{{\mathop  > \limits^{^{{H_0} {\rm{or}} {H_2}}} }}{{\mathop  < \limits_{_{{H_0} {\rm{or}} {H_1}}} }}\frac{1}{{({m_2} - {m_1})}}\ln (\frac{{{\lambda _2}}}{{{\lambda _1}}}) + \frac{{({m_2} + {m_1})}}{2}
\end{cases}
\end{align}
By defining new local decision making thresholds $\gamma_i, i=1, 2, 3$ in  Eq. \ref{Eq:14}, the rules of local detection of two simultaneous events are presented in Eq. \ref{Eq:15}. 
\begin{align}
\label{Eq:14}
\begin{cases}
{\gamma _1} & \buildrel \Delta \over = \frac{1}{{({m_1} - {m_0})}}\ln ({\lambda _1}) + \frac{{({m_1} + {m_0})}}{2}\\
{\gamma _2} & \buildrel \Delta \over = \frac{1}{{({m_2} - {m_0})}}\ln ({\lambda _2}) + \frac{{({m_2} + {m_0})}}{2}\\
{\gamma _3} & \buildrel \Delta \over = \frac{1}{{({m_2} - {m_1})}}(\ln ({\lambda _2}) - \ln ({\lambda _1})) + \frac{{({m_2} + {m_1})}}{2}
\end{cases}
\end{align}
\begin{align}\label{Eq:15}
\begin{cases}
\text{If } & x _i < \gamma _1 \text{ and }  x_i < \gamma _2 \text{ then } H_0 \text{ is selected.}\\
\text{If } & x _i > \gamma _1 \text{ and }  x_i < \gamma _3   \text{ then } H_1 \text{ is selected.}\\
\text{If } & x _i > \gamma _2 \text{ and }  x_i > \gamma _3 \text{ then } H_2 \text{ is selected.}
\end{cases}
\end{align}

\subsection{Probability of Detection Error} \label{Sec:2.2}
In order to find an accurate detection, it is necessary to find the optimal thresholds of decision rules of Eq. \ref{Eq:15}. To this aim, the decision criteria for detection of two simultaneous events and then the probability of detection error ($P_e$) need to be defined. The decision criteria of two simultaneous events are as follow:
\begin{itemize}
\item $P_{D1}$ ($P_{D2}$) is the identical probability of detection, means that \textit{event1} (\textit{event2}) is detected correctly.
\item $P_{F1}$ ($P_{F2}$) is the identical probability of false alarm, means that \textit{event1} (\textit{event2}) is detected while no event is happened.
\item $P_{M1}$ ($P_{M2}$) is the identical probability of misplaced detection, means that \textit{event2} (\textit{event1}) is detected while \textit{event1} (\textit{event2}) is happened. 
\end{itemize}
Assuming that the decisions 0, 1, -1 represent the normal, \textit{event1} and \textit{event2} respectively, the local decision criteria are formulated as: 
\begin{align}
\label{Eq:16}
\begin{cases}
{P_{D1}} & = P({u_i} = 1|{H_1})\\
{P_{D2}} & = P({u_i} =  - 1|{H_2})\\
{P_{F1}} & = P({u_i} = 1 |{H_0})\\
{P_{F2}} & = P({u_i} =  - 1 |{H_0})\\
{P_{M1}} & = P({u_i} =  - 1|{H_1})\\
{P_{M2}} & = P({u_i} = 1|{H_2})
\end{cases}
\end{align}
In a NSN with $n$ sensor nodes, the quality of  final decision at each sensor node in the second layer of the detection algorithm is formulated based on the trinomial distributions. The probability of detection of \textit{event1} ($Q_{D1}$), the probability of detection of \textit{event2} ($Q_{D2}$), the probability of false alarm detection of \textit{event1} ($Q_{F1}$) and the probability of false alarm detection of \textit{event2} ($Q_{F2}$) are as follow:
\begin{align}
\begin{cases}
{Q_{D1}} & = \sum\limits_{j = 0}^{n - i} {\sum\limits_{i = k}^n {\left( \begin{array}{l}
n\\
i,j
\end{array} \right)} \,{P_{D1}}^i{P_{M1}}^j{{(1 - {P_{D1}} - {P_{M1}})}^{n - i - j}}} \\
{Q_{D2}} & = \sum\limits_{j = 0}^{n - i} {\sum\limits_{i = k}^n {\left( \begin{array}{l}
n\\
i,j
\end{array} \right)} \,{P_{D2}}^i{P_{M2}}^j{{(1 - {P_{D2}} - {P_{M2}})}^{n - i - j}}}\\
{Q_{F1}} & = \sum\limits_{j = 0}^{n - i} {\sum\limits_{i = k}^n {\left( \begin{array}{l}
n\\
i,j
\end{array} \right)} \,{P_{F1}}^i{P_{F2}}^j{{(1 - {P_{F1}} - {P_{F2}})}^{n - i - j}}}\\
{Q_{F2}} & = \sum\limits_{j = 0}^{n - i} {\sum\limits_{i = k}^n {\left( \begin{array}{l}
n\\
i,j
\end{array} \right)\,} {P_{F2}}^i{P_{F1}}^j{{(1 - {P_{F1}} - {P_{F2}})}^{n - i - j}}}
\end{cases}
\end{align}
where
\begin{equation}
\left( \begin{array}{l}
n\\
x,y
\end{array} \right)=\dfrac{n!}{x!y!(n-x-y)!}
\end{equation}
The probability of false alarm in the NSN is obtained by:
\begin{equation}
{Q_F} = {Q_{F1}} + {Q_{F2}}
\end{equation}
Based on the Bayesian detection criteria, the probability of error in detection of two events is:
\begin{equation}\label{Eq:27}
{P_e} = {q_0}({Q_F}) + {q_1}(1 - {Q_{D1}}) + {q_2}(1 - {Q_{D2}})
\end{equation}
where $q_0$, $q_1$, $q_2$ are prior probabilities of hypotheses $H_0$, $H_1$ and $H_2$ respectively.  These probabilities are determined based on information about the environment before the experiment is conducted.  According to the assumption of Gaussian distribution for observation $x_i$ in each hypothesis, the local decision criteria are represented by:
\begin{align}
\label{Eq:28}
\begin{cases}
{P_{D1}} = P(x \ge {\gamma _1} \cap x < {\gamma _3}|{H_1} )\\
{P_{D2}} = P(x \ge {\gamma _2} \cap x \ge {\gamma _3}|{H_2} )\\
{P_{F1}} = P(x \ge {\gamma _1} \cap x < {\gamma _3}|{H_0})\\
{P_{F2}} = P(x \ge {\gamma _2} \cap x \ge {\gamma _3}|{H_0})\\
{P_{M1}} = P(x \ge {\gamma _2} \cap x \ge {\gamma _3}|{H_1})\\
{P_{M2}} = P(x \ge {\gamma _1} \cap x < {\gamma _3}|{H_2})
\end{cases}
\end{align}
These equations show the dependency of $P_{D1}$, $P_{D2}$, $P_{F1}$, $P_{F2}$, $P_{M1}$ and $P_{M2}$ on $\gamma _1$, $\gamma _2$ and $\gamma _3$.  Appendix \ref{sec.foo} describes the final equations for calculating the probability of detection error of two simultaneous events $P_e$, which is a function of the thresholds of decision rules $\lambda _1$ and $\lambda _2$.  For the given $q_0$, $q_1$, $q_2$, $n$ and $k$, the $P_e$ is a nonlinear piecewise function. The minimum of $P_e$ and the corresponding thresholds can be found by numerical methods in MATLAB software.  

\subsection{Probability of Detection Error in the Presence of Sensor Faults} \label{Sec:2.3}
In this section, we study the case that there are possibilities of faults $P_f$ in the local decision making of the sensor nodes for detection of two simultaneous events. For this purpose, the probability of detection error in Eq. \ref{Eq:27} is modified to find the optimal thresholds of decision making.  Different types of sensor faults in the procedure of decision making are defined in Table \ref{Table:I}. The $\alpha_i$ , $i=1,\dots,6$ represent the probabilities of different sensor faults in the decision making. The \textit{original} and \textit{reported} decisions are  the decisions of an unfaulty and a faulty sensor respectively.  
\begin{equation}
{P_f} = {\alpha _1} + {\alpha _2} + {\alpha _3} + {\alpha _4} + {\alpha _5} + {\alpha _6}
\end{equation}
\begin{table}[]
\centering
\caption{Probabilities of Different Types of Sensor Fault in Decision Making based on Original and Reported Decisions. }
\label{Table:I}
\begin{tabular}{|l|l|l|l|l|l|l|}
\hline
Decision&${\alpha _1}$ & ${\alpha _2}$ &$ {\alpha _3}$ &$ {\alpha _4}$ &${\alpha _5}$&$ {\alpha _6}$ \\ \hline
Original&	1&	-1&	1&	-1&	0&	0\\ \hline
Reported&	0&	0&	-1	&1	&1	&-1\\ \hline
\end{tabular}
\end{table}
According to the sensor faults, probabilities of first event detection ${\tilde P_{D1}}$, second event detection ${\tilde P_{D2}}$, first event false alarm  ${\tilde P_{F1}}$, second event false alarm ${\tilde P_{F2}}$, first event misplace detection ${\tilde P_{M1}}$, and second event misplace detection ${\tilde P_{M2}}$ are modified in Eq. \ref{Eq:35}. 
\begin{align}
\label{Eq:35}
\begin{cases}
{\tilde P_{D1}} & = {P_{D1}} + {\alpha _4}{P_{M1}} + {\alpha _5}(1 - {P_{D1}} - {P_{M1}}) - ({\alpha _1} + {\alpha _3}){P_{D1}}\\
{\tilde P_{D2}} & = {P_{D2}} + {\alpha _3}{P_{M2}} + {\alpha _6}(1 - {P_{D2}} - {P_{M2}}) - ({\alpha _2} + {\alpha _4}){P_{D2}}\\
{\tilde P_{F1}} & = {P_{F1}} + {\alpha _4}{P_{F2}} + {\alpha _5}(1 - {P_{F1}} - {P_{F2}}) - ({\alpha _1} + {\alpha _3}){P_{F1}}\\
{\tilde P_{F2}} & = {P_{F2}} + {\alpha _3}{P_{F1}} + {\alpha _6}(1 - {P_{F1}} - {P_{F2}}) - ({\alpha _2} + {\alpha _4}){P_{F2}}\\
{\tilde P_{M1}} & = {P_{M1}} + {\alpha _3}{P_{D1}} + {\alpha _6}(1 - {P_{D1}} - {P_{M1}}) - ({\alpha _2} + {\alpha _4}){P_{M1}}\\
{\tilde P_{M2}} & = {P_{M2}} + {\alpha _4}{P_{D2}} + {\alpha _5}(1 - {P_{D2}} - {P_{M2}}) - ({\alpha _1} + {\alpha _3}){P_{M2}}
\end{cases}
\end{align}
The quality of final decisions in a NSN including $n$ sensor nodes are defined as ${\tilde Q_{D1}}$, ${\tilde Q_{D2}}$, ${\tilde Q_{F1}}$, and ${\tilde Q_{F2}}$ where:
\begin{align}
\label{Eq:41}
\begin{cases}
{\tilde Q_{D1}} & = \sum\limits_{j = 0}^{n - i} {\sum\limits_{i = k}^n {\left( \begin{array}{l}
n\\
i,j
\end{array} \right)} {{\tilde P}_{D1}}^i{{\tilde P}_{M1}}^j{{(1 - {{\tilde P}_{D1}} - {{\tilde P}_{M1}})}^{n - i - j}}}\\
{\tilde Q_{D2}} & = \sum\limits_{j = 0}^{n - i} {\sum\limits_{i = k}^n {\left( \begin{array}{l}
n\\
i,j
\end{array} \right)} {{\tilde P}_{D2}}^i{{\tilde P}_{M2}}^j{{(1 - {{\tilde P}_{D2}} - {{\tilde P}_{M2}})}^{n - i - j}}}\\
{\tilde Q_{F1}} & = \sum\limits_{j = 0}^{n - i} {\sum\limits_{i = k}^n {\left( \begin{array}{l}
n\\
i,j
\end{array} \right)} {{\tilde P}_{F1}}^i{{\tilde P}_{F2}}^j{{(1 - {{\tilde P}_{F1}} - {{\tilde P}_{F2}})}^{n - i - j}}}\\
{\tilde Q_{F2}} & = \sum\limits_{j = 0}^{n - i} {\sum\limits_{i = k}^n {\left( \begin{array}{l}
n\\
i,j
\end{array} \right)} {{\tilde P}_{F2}}^i{{\tilde P}_{F1}}^j{{(1 - {{\tilde P}_{F1}} - {{\tilde P}_{F2}})}^{n - i - j}}}
\end{cases}
\end{align}
The probability of false alarm in the NSN is: 
\begin{equation}\label{Eq:45}
{\tilde Q_F} = {\tilde Q_{F1}} + {\tilde Q_{F2}}
\end{equation}
The probability of detection error for two simultaneous events in presence of sensor faults is defined by:
\begin{equation}\label{Eq:46}
{\tilde P_e} = {q_0}({\tilde Q_F}) + {q_1}(1 - {\tilde Q_{D1}}) + {q_2}(1 - {\tilde Q_{D2}})
\end{equation}
Using the obtained equations Eqs. \ref{Eq:35}- \ref{Eq:45} in Appendix \ref{sec.foo}, the optimal thresholds $\lambda _1$ and $\lambda _2$  can be computed for the given $q_0$, $q_1$, $q_2$, $n$ and $k$, and $P_f$ to minimize the probability of detection error of two simultaneous events in presence of sensor faults.  

\section{Illustrative Examples} \label{Sec:3}

To evaluate the accuracy of the proposed method for detection  of two simultaneous events, different experiments are performed in a simulated WSN area with spatial dimension of $20 \times 20$. In the first two experiments, effectiveness  of the detection algorithm for detection of two events with/without sensor faults are examined.  In these experiments, 200 sensors are randomly scattered on the area, where the first event is in a $10 \times 10$ square in the bottom-left corner and the second event is in an $8 \times 8$ square in the upper-right corner.  The prior probability of the normal $q_0$, first event $q_1$, and second event $q_2$ are equal to 0.59, 0.25, and 0.16 respectively.  The other parameters of the events and decision making are chosen as $n=5$, $k=3$, $m_0=0$, $m_1=3$, and $m_2=6$. In the third experiment, we test the effects of different parameters including total number of sensor nodes in WSN, values of $n$ and $k$, probability of sensor faults, prior probabilities and means of sensor observations in the normal and events regions on performance of the decision making and event detection.  
\subsection*{Experiment 1: Detection without Sensor Fault} \label{Sec:3.1}
\begin{figure}[p]
\centering
\includegraphics[scale=.9]{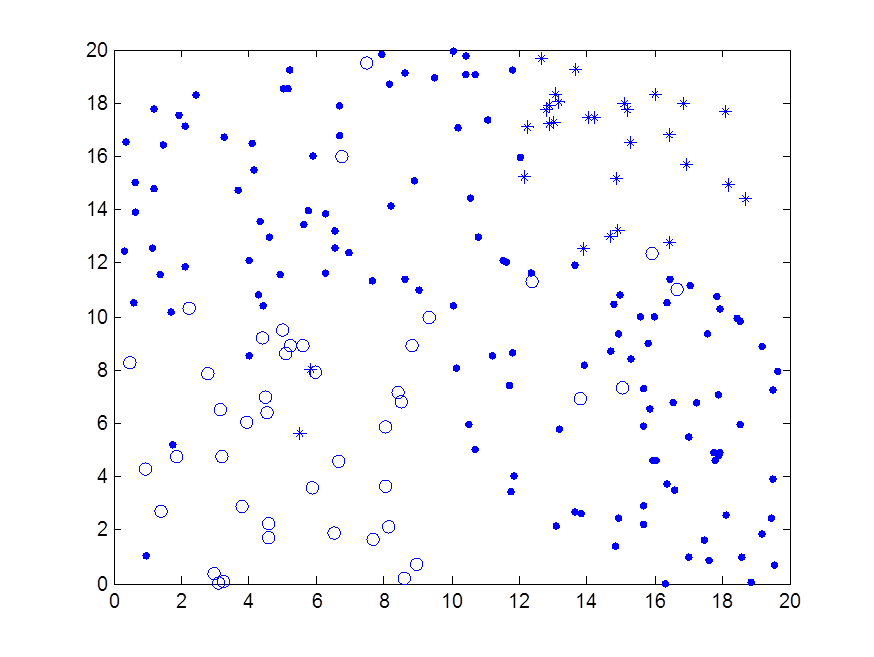}
\vspace{-.5cm}
\caption{Local decisions at the first layer of the detection algorithm with  6.5\% error. The decision is made at the location of each sensor node assuming that there is no fault in the decision making of sensors.}
\label{f4}
\end{figure}

\begin{figure}[p]
\centering
\includegraphics[scale=.9]{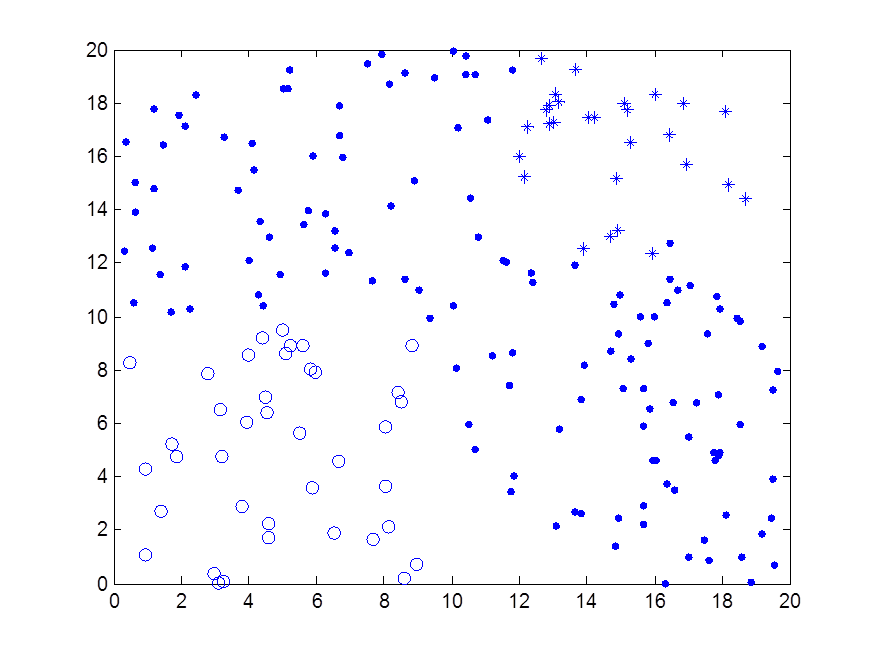}
\vspace{-.5cm}
\caption{Final decisions at the second layer of the detection algorithm with 1.5\% error. The decision is made at the location of each sensor node using the local decisions of the neighboring nodes assuming that there is no fault in the local decision making of sensors.}
\label{f5}
\end{figure}

Based on the chosen  parameters, the optimal thresholds for decision making are found as $\lambda_1=0.9829$, $\lambda_2=1.8496$ which minimize the probability of event detection error in Eq. \ref{Eq:27}. By using these thresholds in decision rules of Eqs. \ref{Eq:14}-\ref{Eq:15}, the result of local detection by each sensor node for two events is shown in Fig. \ref{f4} where the error of detection is 6.5\%.  In this figure ``.'', ``o'' and ``*'' indicate no event, first event and second event at the sensor nodes respectively.  In the second detection layer, decisions are made using the modified $k$ out of $n$ rule as explained in Section \ref{Sec:2}.  Fig. \ref{f5} shows the decision in each sensor node in this layer where the error of decision making is decreased to 1.5\%. It means that only a few number of sensor nodes have detected the events wrongly.
\subsection*{Experiment 2: Detection with Sensor Fault} \label{Sec:3.2}
\begin{figure}[p]
\centering
\includegraphics[scale=.9]{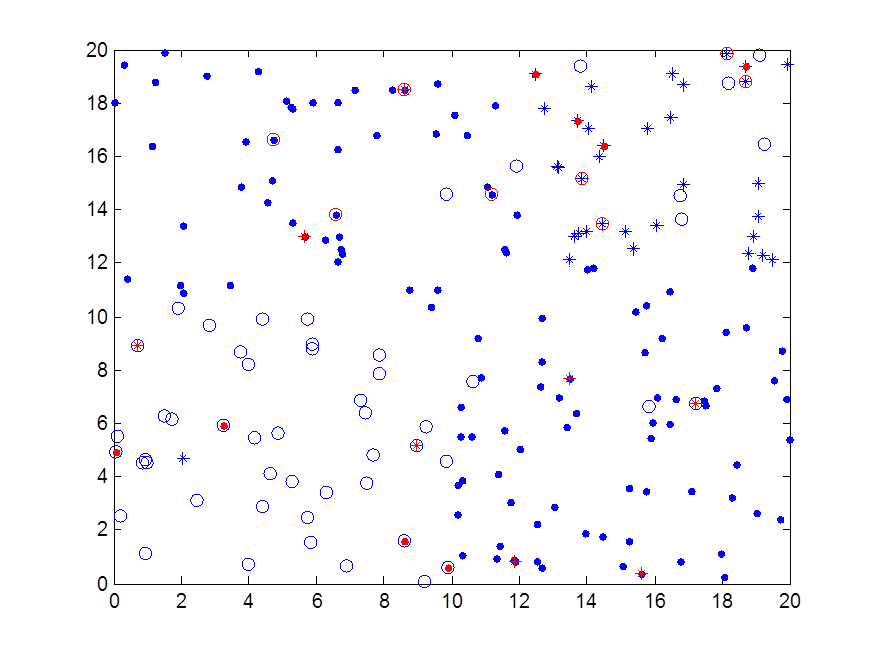}
\vspace{-.5cm}
\caption{Local decisions of 12\% of the sensor nodes are assumed to be faulty which are shown withe red color. Detection error in the first layer has increased to 17.5\%. }
\label{f6}
\end{figure}

\begin{figure}[p]
\centering
\includegraphics[scale=.9]{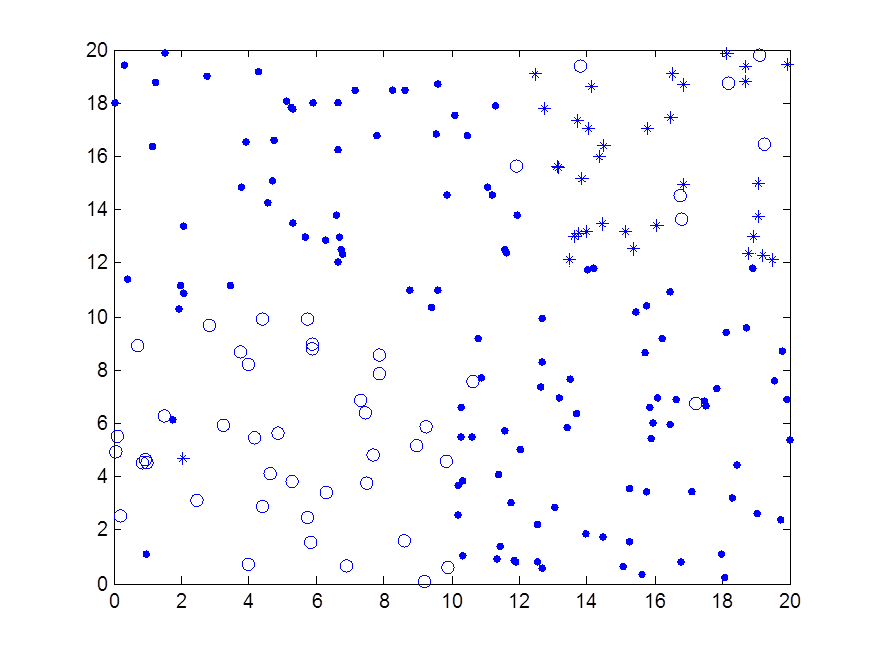}
\vspace{-.5cm}
\caption{Final decisions at the second layer of the detection algorithm where $P_f=0.12$. The detection error is decreased to 11\% using the decisions of neighboring sensors.}
\label{f7}
\end{figure}

Next, we investigate our method in a situation that 12\% of the sensor nodes have faults in their local decision making ($P_f$=0.12). Using numerical method to minimize the probability of detection error (Eq. \ref{Eq:46}), we found that   the optimal $\lambda_1$ and $\lambda_2$ are  0.9504 and 1.7231.  Fig. \ref{f6} shows the local decisions in WSN area while decisions of 12\% random nodes are changed due to the faults. In this figure, faulty sensors are marked by red ``.'', ``o'' and ``*''. In this simulation, by adding the sensor fault, error of local detection of two events reaches to 17.5\%. Then, after final decision making, the error of detection is decreased to 11\%. Results of the detection in the second layer of the algorithm is shown in Fig. \ref{f7}.

\subsection*{Experiment 3: Sensitivity of the Event Detection Algorithm to Parameters}
In this experiment, we test the performance of the detection procedure for different parameters of the network, events and algorithm. In each case of the experiment, the detection algorithm is repeated 50 times, and the sensor observations are randomly generated values. Then, the averages of the following detection errors (in percentage) are compared:

\begin{itemize}
\item LD-BF: the probability of error in the local detection before sensor fault
\item FD-BF: the probability of error in the final detection before sensor fault
\item LD-AF: the probability of error in the local detection after sensor fault  
\item FD-AF: the probability of error in the final detection after sensor fault 
\end{itemize}

First, we examine the performance of the detection in for different values of probability of sensor fault $P_f$.  For each $P_f$, the optimal thresholds computed by the numerical method are found and the averages of the detection errors are compared as in Table \ref{Table:II}. The results show that the proposed method detects two simultaneous events with a high accuracy even in presence of high probability of sensor fault. 
\begin{table}[]
\centering
\caption{Effect of the probability of sensor fault $P_f$ on error of two events detection}
\vspace{.3cm}
\label{Table:II}
\begin{tabular}{|l|l|l|l|l|l|l|}
\hline
$P_f$& LD-BF\% & FD-BF\% & LD-AF\% & FD-AF\% & $\lambda_1$ & $\lambda_2$ \\ \hline
12	&8.20	&3.55	&17.36	&8.01	&0.95	&1.72\\ \hline
24	&8.08	&3.83	&24.11	&13.84	&0.93	&1.64\\ \hline
36	&8.12	&3.73	&29.57	&18.09	&0.92	&1.59\\ \hline
\end{tabular}
\end{table}

The simulation is followed by changing the number of neighboring sensors ($n$ and $k$) for the final decision making. Table \ref{Table:III} indicates that when the number of participated neighbors in the decision making increases, the percentage of the detection error in the second layer decreases. However, using many nodes does not significantly decrease the error. Therefore to reduce the cost of the communication, we can only use a few nodes in the second layer of the detection algorithm without a significant performance degradation.
\begin{table}[]
\centering
\caption{Effect of $n$ and $k$ on error of two events detection}
\vspace{.3cm}
\label{Table:III}
\begin{tabular}{|l|l|l|l|l|l|l|}
\hline
$[n,k]$& LD-BF\% & FD-BF\% & LD-AF\% & FD-AF\% & $\lambda_1$ & $\lambda_2$ \\ \hline
[3,2]	&7.9	&4.2	&16.6&	9.4	&1.2	&2.2\\ \hline
[5,3]	&8.1	&3.7	&16.9&	7.8&	0.9	&1.7\\ \hline
[7,4]	&8.6	&3.8	&17.7&	7.2&	0.8	&1.5\\ \hline
[9,5]	&8.3	&4.0	&17.5&	7.1	&0.7&	1.3\\ \hline
\end{tabular}
\end{table}

In Table \ref{Table:IV}, the effect of number of sensors $N$ distributed in the area has been studied. The results show that by employing large number of sensor in the interested area, probability of detection error is decreased after final processing. But our method works well even if there are not too many sensors in the area.
\begin{table}[]
\centering
\caption{Effect on number of sensor nodes $N$ on error of two events detection}
\vspace{.3cm}
\label{Table:IV}
\begin{tabular}{|l|l|l|l|l|l|l|}
\hline
$N$& LD-BF\% & FD-BF\% & LD-AF\% & FD-AF\% & $\lambda_1$ & $\lambda_2$ \\ \hline
200	&8.1	&3.7	&16.9	&7.8	&0.95	&1.7\\ \hline
400	&8.4	&2.9	&17.3	&6.7	&0.95	&1.7\\ \hline
700	&8.3	&2.1	&17.3	&6.4	&0.95	&1.7\\ \hline
1000	&8.1	&1.9&	17.1	&5.9	&0.95	&1.7\\ \hline
\end{tabular}
\end{table}
Another test is accomplished to study the effect of mean of the sensor observations in normal, first event and second event regions.  Obtained results in Table \ref{Table:V} show that the proposed method works well even if the difference between mean of the observations in different area are not too much.
\begin{table}[]
\centering
\caption{Effect of mean of the sensor observations ($m_0$, $m_1$ and $m_2$) on error of two events detection}
\vspace{.3cm}
\label{Table:V}
\begin{tabular}{|l|l|l|l|l|l|l|}
\hline
$[m_0, m_1, m_2]$& LD-BF\% & FD-BF\% & LD-AF\% & FD-AF\% & $\lambda_1$ & $\lambda_2$ \\ \hline
[0,3,6]	&18.1	&3.7	&16.9	&7.8	&0.9	&1.7 \\ \hline
[0,4,9]	&2.3	&2.6	&12.4	&5.3&	1.0	&2.6 \\ \hline
[-6,-3,-1]	&12.5	&5.5	&20.3	&11	&0.7	&0.9 \\ \hline
\end{tabular}
\end{table}

In addition, we compare the effect of different prior probability of the hypothesis which are equivalent with prior information about the area of the events. The obtained results in Table \ref{Table:VI} indicate that the probability of detection error are small even for the case that the difference between prior probabilities of normal situation and events hypothesis are not huge, means that we can correctly detect those events whose statistical distributions are similar to each other.
\begin{table}[]
\centering
\caption{Effect of prior probabilities ($q_0$, $q_1$ and $q_2$) on error of two events detection}
\vspace{.3cm}
\label{Table:VI}
\begin{tabular}{|l|l|l|l|l|l|l|}
\hline
$[q_0, q_1, q_2]$                                           & LD-BF\% & FD-BF\% & LD-AF\% & FD-AF\% & $\lambda_1$ & $\lambda_2$ \\ \hline
[0.875,0.0625, 0.0625]                                       & 5       & 1.9     & 11.4    & 6.4     & 1.7        & 3.5        \\ \hline
[0.82, 0.09, 0.09]                                           & 5.6     & 2.2     & 13.1    & 7.4     & 1.7        & 2.7        \\ \hline
\begin{tabular}[c]{@{}l@{}}[0.59,0.25, 0.16]\end{tabular} & 8.1     & 3.7     & 16.9    & 7.8     & 0.9        & 1.7        \\ \hline
\begin{tabular}[c]{@{}l@{}}[0.5,0.25, 0.25]\end{tabular}  & 9       & 4       & 18.1    & 8       & 0.8        & 1.2        \\ \hline
\end{tabular}
\end{table}

\section{Conclusion}
Detection of simultaneous events in a WSN area is a significant problem which has many applications in industry. This paper is a new attempt on distributed event detection where only information of the neighboring sensor nodes are used for decision making.  To this aim, we introduced a new set of decision criteria to formulate the probability of detection error based on Bayesian criterion and likelihood ratio test.  By minimizing the the probability of detection error, optimal thresholds for decision making are obtained.  In addition, our proposed method is expanded to a fault-tolerant detection procedure if the sensor faults corrupt the decisions in some nodes. Simulations show the accuracy of the proposed method for various values of probabilities of sensor faults, number of neighbor sensors, prior probabilities of events, number of sensor nodes and means of data distributions in event regions.

\appendix\section{Appendix}\label{sec.foo}
Since observations have a Gaussian distribution in each hypothesis, so decision criteria proposed in Eq.  \ref{Eq:28} can be represent in a simplified format with five different cases of decision thresholds.
\begin{itemize}
\item Case 1: ${\gamma _1} < {\gamma _2} < {\gamma _3}$
\begin{align*}
{P_{D1}} & = P({\gamma _1} \le x < {\gamma _3}|{H_1}) = \int\limits_{{\gamma _1} - {m_1}}^{{\gamma _3} - {m_1}} {\frac{1}{{\sqrt {2\pi } }}} exp(\frac{{ - {x^2}}}{2})dx\\
{P_{D2}} & = P(x \ge {\gamma _3}|{H_2}) = \int\limits_{{\gamma _3} - {m_2}}^\infty  {\frac{1}{{\sqrt {2\pi } }}} exp(\frac{{ - {x^2}}}{2})dx\\
{P_{F1}} & = P({\gamma _1} \le x < {\gamma _3}|{H_0}) = \int\limits_{{\gamma _1} - {m_0}}^{{\gamma _3} - {m_0}} {\frac{1}{{\sqrt {2\pi } }}} exp(\frac{{ - {x^2}}}{2})dx\\
{P_{F2}} & = P(x \ge {\gamma _3}|{H_0}) = \int\limits_{{\gamma _3} - {m_0}}^\infty  {\frac{1}{{\sqrt {2\pi } }}} exp(\frac{{ - {x^2}}}{2})dx\\
{P_{M1}} & = P(x \ge {\gamma _3}|{H_1}) = \int\limits_{{\gamma _3} - {m_1}}^\infty  {\frac{1}{{\sqrt {2\pi } }}} exp(\frac{{ - {x^2}}}{2})dx\\
{P_{M2}} & = P({\gamma _1} \le x < {\gamma _3}|{H_2}) = \int\limits_{{\gamma _1} - {m_2}}^{{\gamma _3} - {m_2}} {\frac{1}{{\sqrt {2\pi } }}} exp(\frac{{ - {x^2}}}{2})dx
\end{align*}

\item Case 2: ${\gamma _1} < {\gamma _3} < {\gamma _2} $
\begin{align*}
{P_{D1}} & = P({\gamma _1} \le x < {\gamma _3}|{H_1}) = \int\limits_{{\gamma _1} - {m_1}}^{{\gamma _3} - {m_1}} {\frac{1}{{\sqrt {2\pi } }}} exp(\frac{{ - {x^2}}}{2})dx\\
{P_{D2}} & = P(x \ge {\gamma _2}|{H_2}) = \int\limits_{{\gamma _2} - {m_2}}^\infty  {\frac{1}{{\sqrt {2\pi } }}} exp(\frac{{ - {x^2}}}{2})dx\\
{P_{F1}} & = P({\gamma _1} \le x < {\gamma _3}|{H_0}) = \int\limits_{{\gamma _1} - {m_0}}^{{\gamma _3} - {m_0}} {\frac{1}{{\sqrt {2\pi } }}} exp(\frac{{ - {x^2}}}{2})dx\\
{P_{F2}} & = P(x \ge {\gamma _2}|{H_0}) = \int\limits_{{\gamma _2} - {m_0}}^\infty  {\frac{1}{{\sqrt {2\pi } }}} exp(\frac{{ - {x^2}}}{2})dx\\
{P_{M1}} & = P(x \ge {\gamma _2}|{H_1}) = \int\limits_{{\gamma _2} - {m_1}}^\infty  {\frac{1}{{\sqrt {2\pi } }}} exp(\frac{{ - {x^2}}}{2})dx\\
{P_{M2}} & = P({\gamma _1} \le x < {\gamma _3}|{H_2}) = \int\limits_{{\gamma _1} - {m_2}}^{{\gamma _3} - {m_2}} {\frac{1}{{\sqrt {2\pi } }}} exp(\frac{{ - {x^2}}}{2})dx
\end{align*}

\item Case 3: ${\gamma _2} < {\gamma _1} < {\gamma _3}$

\begin{align*}
{P_{D1}} & = P({\gamma _1} \le x < {\gamma _3}|{H_1}) = \int\limits_{{\gamma _1} - {m_1}}^{{\gamma _3} - {m_1}} {\frac{1}{{\sqrt {2\pi } }}} exp(\frac{{ - {x^2}}}{2})dx\\
{P_{D2}} & = P(x \ge {\gamma _3}|{H_2}) = \int\limits_{{\gamma _3} - {m_2}}^\infty  {\frac{1}{{\sqrt {2\pi } }}} exp(\frac{{ - {x^2}}}{2})dx\\
{P_{F1}} & = P({\gamma _1} \le x < {\gamma _3}|{H_0}) = \int\limits_{{\gamma _1} - {m_0}}^{{\gamma _3} - {m_0}} {\frac{1}{{\sqrt {2\pi } }}} exp(\frac{{ - {x^2}}}{2})dx\\
{P_{F2}} & = P(x \ge {\gamma _3}|{H_0}) = \int\limits_{{\gamma _3} - {m_0}}^\infty  {\frac{1}{{\sqrt {2\pi } }}} exp(\frac{{ - {x^2}}}{2})dx \\
{P_{M1}} & = P(x \ge {\gamma _3}|{H_1}) = \int\limits_{{\gamma _3} - {m_1}}^\infty  {\frac{1}{{\sqrt {2\pi } }}} exp(\frac{{ - {x^2}}}{2})dx\\
{P_{M2}} & = P({\gamma _1} \le x < {\gamma _3}|{H_2}) = \int\limits_{{\gamma _1} - {m_2}}^{{\gamma _3} - {m_2}} {\frac{1}{{\sqrt {2\pi } }}} exp(\frac{{ - {x^2}}}{2})dx
\end{align*}

\item Case 4: ${\gamma _2} < {\gamma _3} < {\gamma _1}$

\begin{align*}
{P_{D1}} & = 0\\
{P_{D2}} & = P(x \ge {\gamma _3}|{H_2}) = \int\limits_{{\gamma _3} - {m_2}}^\infty  {\frac{1}{{\sqrt {2\pi } }}} exp(\frac{{ - {x^2}}}{2})dx\\
{P_{F1}} & = P(\emptyset |{H_0}) = 0\\
{P_{F2}} & = P(x \ge {\gamma _3}|{H_0}) = \int\limits_{{\gamma _3} - {m_0}}^\infty  {\frac{1}{{\sqrt {2\pi } }}} exp(\frac{{ - {x^2}}}{2})dx\\
{P_{M1}} & = P(x \ge {\gamma _3}|{H_1}) = \int\limits_{{\gamma _3} - {m_2}}^\infty  {\frac{1}{{\sqrt {2\pi } }}} exp(\frac{{ - {x^2}}}{2})dx\\
{P_{M2}} & = 0
\end{align*}

\item Case 5: ${\gamma _3} < {\gamma _1} < {\gamma _2}$  and  ${\gamma _3} < {\gamma _2} < {\gamma _1}$

\begin{align*}
{P_{D1}} & = 0\\
{P_{D2}} & = P(x \ge {\gamma _2}|{H_2}) = \int\limits_{{\gamma _2} - {m_2}}^\infty  {\frac{1}{{\sqrt {2\pi } }}} exp(\frac{{ - {x^2}}}{2})dx\\
{P_{F2}} & = P(\emptyset |{H_0}) = 0\\
{P_{F2}} & = P(x \ge {\gamma _2}|{H_0}) = \int\limits_{{\gamma _2} - {m_0}}^\infty  {\frac{1}{{\sqrt {2\pi } }}} exp(\frac{{ - {x^2}}}{2})dx\\
{P_{M1}} & = P(x \ge {\gamma _2}|{H_1}) = \int\limits_{{\gamma _2} - {m_2}}^\infty  {\frac{1}{{\sqrt {2\pi } }}} exp(\frac{{ - {x^2}}}{2})dx\\
{P_{M2}} & = 0
\end{align*}

\end{itemize}

\bibliography{myref}
\bibliographystyle{IEEEtrans}

\end{document}